\documentclass[aps,prl,twocolumn,superscriptaddress,nofootinbib]{revtex4}

\usepackage[dvipsnames]{xcolor}
\usepackage{empheq}
\usepackage{gensymb}
\usepackage{graphicx}
\usepackage{adjustbox}
\usepackage{dcolumn}
\usepackage{bm}
\usepackage{hyperref}
\usepackage[mathlines]{lineno}

\newcommand{\ptmu}{$P_{T}$}
\newcommand{\pparmu}{$P_{||}$}

\hypersetup{
    pdfnewwindow=true,      
    colorlinks=true,       
    linkcolor=blue,          
    citecolor=blue,        
    filecolor=blue,      
    urlcolor=blue           
}
\newcommand{\phref}[2]{\href{#2}{#1}}

\let\oldalign\align
\let\oldendalign\endalign

\renewenvironment{align}
  {\linenomathNonumbers\oldalign}
  {\oldendalign\endlinenomath}

\let\oldempheq\empheq
\let\oldendempheq\endempheq

\renewenvironment{empheq}
  {\linenomathNonumbers\oldempheq}
  {\oldendempheq\endlinenomath}

\begin{document}
\title{Simultaneous measurement of $\nu_\mu$ quasielastic-like cross sections on CH, C, H$_2$O, Fe, and Pb as a function of muon kinematics at MINERvA}

\newcommand{\Rutgers}{Rutgers, The State University of New Jersey, Piscataway, New Jersey 08854, USA}
\newcommand{\Hampton}{Hampton University, Dept. of Physics, Hampton, VA 23668, USA}
\newcommand{\Dortmund}{Institute of Physics, Dortmund University, 44221, Germany }
\newcommand{\Otterbein}{Department of Physics, Otterbein University, 1 South Grove Street, Westerville, OH, 43081 USA}
\newcommand{\JMU}{James Madison University, Harrisonburg, Virginia 22807, USA}
\newcommand{\Florida}{University of Florida, Department of Physics, Gainesville, FL 32611}
\newcommand{\UCIrvine}{Department of Physics and Astronomy, University of California, Irvine, Irvine, California 92697-4575, USA}
\newcommand{\CBPF}{Centro Brasileiro de Pesquisas F\'{i}sicas, Rua Dr. Xavier Sigaud 150, Urca, Rio de Janeiro, Rio de Janeiro, 22290-180, Brazil}
\newcommand{\PUCP}{Secci\'{o}n F\'{i}sica, Departamento de Ciencias, Pontificia Universidad Cat\'{o}lica del Per\'{u}, Apartado 1761, Lima, Per\'{u}}
\newcommand{\INRM}{Institute for Nuclear Research of the Russian Academy of Sciences, 117312 Moscow, Russia}
\newcommand{\Jlab}{Jefferson Lab, 12000 Jefferson Avenue, Newport News, VA 23606, USA}
\newcommand{\Pittsburgh}{Department of Physics and Astronomy, University of Pittsburgh, Pittsburgh, Pennsylvania 15260, USA}
\newcommand{\Guanajuato}{Campus Le\'{o}n y Campus Guanajuato, Universidad de Guanajuato, Lascurain de Retana No. 5, Colonia Centro, Guanajuato 36000, Guanajuato M\'{e}xico.}
\newcommand{\Athens}{Department of Physics, University of Athens, GR-15771 Athens, Greece}
\newcommand{\Tufts}{Physics Department, Tufts University, Medford, Massachusetts 02155, USA}
\newcommand{\WM}{Department of Physics, William \& Mary, Williamsburg, Virginia 23187, USA}
\newcommand{\FNAL}{Fermi National Accelerator Laboratory, Batavia, Illinois 60510, USA}
\newcommand{\Purdue}{Department of Chemistry and Physics, Purdue University Calumet, Hammond, Indiana 46323, USA}
\newcommand{\MCLA}{Massachusetts College of Liberal Arts, 375 Church Street, North Adams, MA 01247}
\newcommand{\UMD}{Department of Physics, University of Minnesota -- Duluth, Duluth, Minnesota 55812, USA}
\newcommand{\Northwestern}{Northwestern University, Evanston, Illinois 60208}
\newcommand{\UNI}{Facultad de Ciencias, Universidad Nacional de Ingenier\'{i}a, Apartado 31139, Lima, Per\'{u}}
\newcommand{\Rochester}{Department of Physics and Astronomy, University of Rochester, Rochester, New York 14627 USA}
\newcommand{\Austin}{Department of Physics, University of Texas, 1 University Station, Austin, Texas 78712, USA}
\newcommand{\USM}{Departamento de F\'{i}sica, Universidad T\'{e}cnica Federico Santa Mar\'{i}a, Avenida Espa\~{n}a 1680 Casilla 110-V, Valpara\'{i}so, Chile}
\newcommand{\Geneva}{University of Geneva, 1211 Geneva 4, Switzerland}
\newcommand{\Chicago}{Enrico Fermi Institute, University of Chicago, Chicago, IL 60637 USA}
\newcommand{\hired}{}
\newcommand{\OregonState}{Department of Physics, Oregon State University, Corvallis, Oregon 97331, USA}
\newcommand{\oxford}{Oxford University, Department of Physics, Oxford, OX1 3PJ United Kingdom}
\newcommand{\umiss}{University of Mississippi, Oxford, Mississippi 38677, USA}
\newcommand{\upenn}{Department of Physics and Astronomy, University of Pennsylvania, Philadelphia, PA 19104}
\newcommand{\AMU}{Department of Physics, Aligarh Muslim University, Aligarh, Uttar Pradesh 202002, India}
\newcommand{\wroclaw}{University of Wroclaw, plac Uniwersytecki 1, 50-137 Wroa\l{}aw, Poland}
\newcommand{\Mohali}{Department of Physical Sciences, IISER Mohali, Knowledge City, SAS Nagar, Mohali - 140306, Punjab, India}
\newcommand{\CINVESTAV}{Departamento de Fisica Col. San Pedro Zacatenco, 07360 Mexico, DF, Av. Instituto PolitÃ©cnico Nacional, Mexico}
\newcommand{\york}{York University, Department of Physics and Astronomy, Toronto, Ontario, M3J 1P3 Canada}
\newcommand{\ND}{Department of Physics, University of Notre Dame, Notre Dame, Indiana 46556, USA}
\newcommand{\ICL}{The Blackett Laboratory,  Imperial College London,  London SW7 2BW, United Kingdom}
\newcommand{\warwick}{Department of Physics, University of Warwick, Coventry, CV4 7AL, UK}
\newcommand{\qmul}{G O Jones Building, Queen Mary University of London, 327 Mile End Road, London E1 4NS, UK}

\newcommand{\kleykampThanks}{now at Department of Physics and Astronomy, University of Mississippi, Oxford, MS 38677}
\newcommand{\amitbashyalThanks}{Now at  High Energy Physics/Center for Computational Excellence Department, Argonne National Lab, 9700 S Cass Ave, Lemont, IL 60439}
\newcommand{\ricfregianThanks}{now at Department of Physics and Astronomy, University of California at Davis, Davis, CA 95616, USA}
\newcommand{\mateusfcarneiroThanks}{Now at Brookhaven National Laboratory, Upton, New York 11973-5000, USA}
\newcommand{\finerThanks}{Now at Los Alamos National Laboratory, Los Alamos, New Mexico 87545, USA}
\newcommand{\bamThanks}{Now at University of Minnesota, Minneapolis, Minnesota 55455, USA}
\newcommand{\byaeggyThanks}{Now at Department of Physics, University of Cincinnati,  Cincinnati, Ohio 45221, USA}
\newcommand{\mascencioThanks}{Now at Iowa State University, Ames, IA 50011, USA}
\newcommand{\anfilkinsThanks}{now at Syracuse University, Syracuse, NY 13244, USA}

%
%
%
%

\author{J.~Kleykamp}\thanks{\kleykampThanks}  \affiliation{\Rochester}
\author{S.~Akhter}                        \affiliation{\AMU}
\author{Z.~~Ahmad~Dar}                    \affiliation{\WM}  \affiliation{\AMU}
\author{V.~Ansari}                        \affiliation{\AMU}
\author{M.~V.~Ascencio}\thanks{\mascencioThanks}  \affiliation{\PUCP}
\author{M.~Sajjad~Athar}                  \affiliation{\AMU}
\author{A.~Bashyal}\thanks{\amitbashyalThanks}  \affiliation{\OregonState}
\author{A.~Bercellie}                     \affiliation{\Rochester}
\author{M.~Betancourt}                    \affiliation{\FNAL}
\author{A.~Bodek}                         \affiliation{\Rochester}
\author{J.~L.~Bonilla}                    \affiliation{\Guanajuato}
\author{A.~Bravar}                        \affiliation{\Geneva}
\author{H.~Budd}                          \affiliation{\Rochester}
\author{G.~Caceres}\thanks{\ricfregianThanks}  \affiliation{\CBPF}
\author{T.~Cai}                           \affiliation{\Rochester} \affiliation{\york}
\author{M.F.~Carneiro}\thanks{\mateusfcarneiroThanks}  \affiliation{\OregonState}  \affiliation{\CBPF}
\author{G.A.~D\'{i}az~}                   \affiliation{\Rochester}
\author{H.~da~Motta}                      \affiliation{\CBPF}
\author{S.A.~Dytman}                      \affiliation{\Pittsburgh}
\author{J.~Felix}                         \affiliation{\Guanajuato}
\author{L.~Fields}                        \affiliation{\ND}
\author{A.~Filkins}\thanks{\anfilkinsThanks}  \affiliation{\WM}
\author{R.~Fine}\thanks{\finerThanks}     \affiliation{\Rochester}
\author{A.M.~Gago}                        \affiliation{\PUCP}
\author{H.~Gallagher}                     \affiliation{\Tufts}
\author{S.M.~Gilligan}                    \affiliation{\OregonState}
\author{R.~Gran}                          \affiliation{\UMD}
\author{E.Granados}                       \affiliation{\Guanajuato}
\author{D.A.~Harris}                      \affiliation{\york}  \affiliation{\FNAL}
\author{S.~Henry}                         \affiliation{\Rochester}
\author{D.~Jena}                          \affiliation{\FNAL}
\author{S.~Jena}                          \affiliation{\Mohali}
\author{A.~Klustov\'{a}}                  \affiliation{\ICL}
\author{M.~Kordosky}                      \affiliation{\WM}
\author{D.~Last}                          \affiliation{\upenn}
\author{A.~Lozano}                        \affiliation{\CBPF}
\author{X.-G.~Lu}                         \affiliation{\warwick}  \affiliation{\oxford}
\author{E.~Maher}                         \affiliation{\MCLA}
\author{S.~Manly}                         \affiliation{\Rochester}
\author{W.A.~Mann}                        \affiliation{\Tufts}
\author{C.~Mauger}                        \affiliation{\upenn}
\author{K.S.~McFarland}                   \affiliation{\Rochester}
\author{B.~Messerly}\thanks{\bamThanks}   \affiliation{\Pittsburgh}
\author{J.~Miller}                        \affiliation{\USM}
\author{O.~Moreno}                        \affiliation{\WM}  \affiliation{\Guanajuato}
\author{J.G.~Morf\'{i}n}                  \affiliation{\FNAL}
\author{D.~Naples}                        \affiliation{\Pittsburgh}
\author{J.K.~Nelson}                      \affiliation{\WM}
\author{C.~Nguyen}                        \affiliation{\Florida}
\author{A.~Olivier}                       \affiliation{\Rochester}
\author{V.~Paolone}                       \affiliation{\Pittsburgh}
\author{G.N.~Perdue}                      \affiliation{\FNAL}  \affiliation{\Rochester}
\author{K.-J.~Plows}                      \affiliation{\oxford}
\author{M.A.~Ram\'{i}rez}                 \affiliation{\upenn}  \affiliation{\Guanajuato}
\author{R.D.~Ransome}                     \affiliation{\Rutgers}
\author{H.~Ray}                           \affiliation{\Florida}
\author{D.~Ruterbories}                   \affiliation{\Rochester}
\author{H.~Schellman}                     \affiliation{\OregonState}
\author{C.J.~Solano~Salinas}              \affiliation{\UNI}
\author{H.~Su}                            \affiliation{\Pittsburgh}
\author{M.~Sultana}                       \affiliation{\Rochester}
\author{V.S.~Syrotenko}                   \affiliation{\Tufts}
\author{E.~Valencia}                      \affiliation{\WM}  \affiliation{\Guanajuato}
\author{N.H.~Vaughan}                     \affiliation{\OregonState}
\author{A.V.~Waldron}                     \affiliation{\qmul}  \affiliation{\ICL}
\author{C.~Wret}                          \affiliation{\Rochester}
\author{B.~Yaeggy}\thanks{\byaeggyThanks}  \affiliation{\USM}
\author{L.~Zazueta}      \affiliation{\WM}


\collaboration{The MINER$\nu$A Collaboration}\ \noaffiliation
\date{\today}

\begin{abstract} 
This paper presents the first simultaneous measurement of the quasielastic-like neutrino-nucleus cross sections on C, water, Fe, Pb and scintillator (hydrocarbon or CH) as a function of longitudinal and transverse muon momentum.   
The ratio of cross sections per nucleon between Pb and CH is always above unity and has a characteristic shape as a function of 
transverse muon momentum that evolves slowly as a function of longitudinal muon momentum. The ratio is constant versus longitudinal momentum within uncertainties above a longitudinal momentum of 4.5GeV/c.  
The cross section ratios to CH for C, water, and Fe remain roughly constant with increasing longitudinal momentum, and the ratios between water or C to CH do not have any significant deviation from unity. 
Both the overall cross section level and the shape for Pb and Fe as a function of transverse muon momentum are not reproduced by current neutrino event generators.
These measurements provide a direct test of nuclear effects in quasielastic-like interactions, which are major contributors to long-baseline neutrino oscillation data samples.  
\end{abstract}

\maketitle

The charged-current quasielastic neutrino interaction ({\em i.e.} $\nu_{\mu}n\rightarrow\mu^{-}p$) contributes the majority of selected signal interactions in current accelerator-based neutrino oscillation experiments~\cite{Abe:2018wpn, Abe:2017vif, Acero:2019ksn, NOvA:2018gge, Acciarri:2015uup, Abe:2015zbg}.  
Because the interaction's final state is simple, the lepton flavor is easily identified. The neutrino energy may be estimated assuming two-body kinematics where the target is assumed to be a neutron at rest.  However, for those nuclei in use in oscillation experiments 
these assumptions can bias neutrino energy reconstruction because of finite initial neutron momentum inside the nucleus~\cite{NuSTEC:2017hzk}.  
In addition, quasielastic scattering can be mimicked by other processes;  
for example, when final-state particles 
are absorbed in the nucleus.
These biases are already significant in current experiments~\cite{Abe:2018wpn, Abe:2017vif, Acero:2019ksn, NOvA:2018gge}
and risk becoming dominant uncertainties in the future, for example in
DUNE~\cite{Acciarri:2015uup} and Hyper-Kamiokande~\cite{Abe:2015zbg}.  Oscillation experiments  
may also use different target nuclei for their near detector than for their far detector~\cite{Abe:2018wpn}, so direct measurements on different nuclei provide insight on biases that might be introduced by that choice.

The MINERvA experiment published a measurement of quasielastic-like cross sections on a variety of nuclei at a mean neutrino energy of 3~GeV~\cite{Betancourt:2017uso},  
using interactions where both a final-state muon and proton were identified. This letter describes a measurement made with a data set that is over twenty times as large 
due to the following factors: the mean neutrino energy is higher by a factor of two, the integrated number of protons on target is larger by a factor of four, and the requirement for a final-state proton has been removed. 
The increased statistics allow a more detailed probe of 
this process.  

MINERvA recently measured charged-current charged-pion production on different nuclei~\cite{MINERvA:2022djk}.  Using the same neutrino beam as used here, MINERvA found that the ratio of pions produced on Fe or Pb compared to scintillator is lower than predicted by current models.  This has implications not only for the background, but also for the signal in this analysis.  Neutrino interactions in which pions are produced but absorbed in the nucleus can be quasielastic-like, and are thus included as a signal process.  In addition, due to the possibility of interactions with nucleon pairs, the quasielastic-like definition allows any number of protons and neutrons in the final state.  

The MINERvA detector~\cite{Aliaga:2013uqz} consists of a nuclear target region of several thin passive targets interspersed with 1.7~cm-thick active scintillator planes, followed by a scintillator-only region followed by electromagnetic and hadronic calorimetry.  The MINOS near detector~\cite{Michael:2008bc}, located 2~m downstream of MINERvA, measures the charge and momentum of final-state muons.  MINERvA's targets include regions made up of C, Fe, Pb, and water.  The solid targets are configured in such a way that the total amount of passive material a particle traverses between the start of the interaction and the scintillator-only region (in $g/cm^2$) is approximately the same. 
The water target is a flattened circular neoprene balloon that is between 17-24~cm thick in the beam direction.
The detector is modeled using a hit-level Geant4-based simulation overlaid with random beam data to simulate beam-related accidental activity.  The simulation includes the time dependence of both the proton beam intensity and the configuration of the water target.

The NuMI beam is produced by a $120$-GeV proton beam incident on a two-interaction-length graphite target followed by two parabolic focusing horns and a $675$-m decay pipe and 200~m of earth to shield the tertiary muons. For these data the horn currents are set to focus positively-charged pions, creating a neutrino-dominated broad-band beam with a peak energy of $6.5$~GeV.

The beam line is modeled with a Geant4-based~\cite{Agostinelli:2002hh,Allison:2006ve}  simulation (g4numi~\cite{Aliaga:2016oaz} version~6, built against Geant version v.9.4.p2). There are known discrepancies between measurements and Geant4 predictions of pion production from proton-on-carbon interactions relevant to NuMI flux predictions~\cite{Aliaga:2016oaz}. MINERvA corrects these predictions using hadron-production data. In addition, measurements of neutrino-electron ($\nu-e$) scattering~\cite{MINERvA:2019mkf} and interactions with low recoil energy~\cite{MINERvA:2021mpk} are used to constrain the flux prediction.  This analysis uses data that correspond to $10.61\times 10^{20}$ protons on target (POT), where the first (second) half of the exposure was with the water target empty (full) to allow the non-water background interactions to be measured directly and subtracted from the full target sample.  
 
The GENIE 2.12.6 event generator~\cite{Andreopoulos:2009rq} is used to simulate neutrino interactions on nuclei.  For quasielastic scattering on nucleons the Llewellyn-Smith formalism is used~\cite{Llewellyn:1972}.  Nuclear effects are incorporated by using a Bodek-Ritchie high momentum tail~\cite{Bodek:1981wr} in the Fermi momentum distribution of the initial-state nucleons.  
The default GENIE interaction model is adjusted to match previous MINERvA data via a GENIE tune v1 (MnvGENIEv1), which includes three additional modifications.  First, the Valencia Random Phase Approximation (RPA) correction, considered as a \lq\lq weak nuclear screening"~\cite{Nieves:2004wx,Gran:2017psn} for a Fermi gas~\cite{Martini:2009,Nieves:2017lij}, is added as a function of neutrino energy and three-momentum transfer.  Second, the prediction for multi-nucleon scattering given by the Valencia model~\cite{Nieves:2011pp,Gran:2013kda,Schwehr:2016pvn} in GENIE 2.12.6 is added and modified with an empirical fit~\cite{Rodrigues:2015hik} based on previous MINERvA data on CH. The modification increases the integrated 2-particle 2-hole (2p2h) interaction rate by 49\%.  This same fractional increase per proton-neutron pair is applied for all nuclei.  
Finally, non-resonant pion production is reduced by $57\%$ to agree with a fit to measurements  
on deuterium~\cite{Rodrigues:2016xjj}. 


Interactions are selected by requiring a muon candidate that originates in the MINERvA detector and is reconstructed in the MINOS near detector.  There is no minimum number requirement for proton tracks, and there must be no electron candidates resulting from the pion to muon to electron decay chain (\lq\lq Michel"s) near the interaction vertex or any track endpoint.  Backgrounds and efficiencies are determined separately for 
the samples with and without the identified proton tracks.
To further reject interactions containing charged pions,
any non-muon reconstructed track is required to satisfy proton identification cuts based on the energy deposition pattern.  To remove interactions with neutral pions, a cut is made requiring no more than one isolated cluster of energy in the detector.  

The muon momentum is found by the addition of the momentum determined by range inside the MINERvA detector plus the momentum determined by range or curvature inside MINOS~\cite{MINERvA:2018hqn}.  The muon angle is measured in the MINERvA detector. 
To address the MINOS acceptance, only interactions with muons reconstructed within $17^{o}$ of the neutrino beam and with momenta above 1.5~GeV/c and below 40~GeV/c are retained.  The cross sections we report are defined as any interaction with a muon in the final state, where the muon has an angle of no more than $17^o$ and a momentum between 2 and 20~GeV/c.  For these interactions any number of nucleons is allowed, but no photons above 10~MeV (to accommodate nuclear excitations) and no mesons.
  
\begin{figure}[h]
    \centering
\includegraphics[width=\linewidth]
{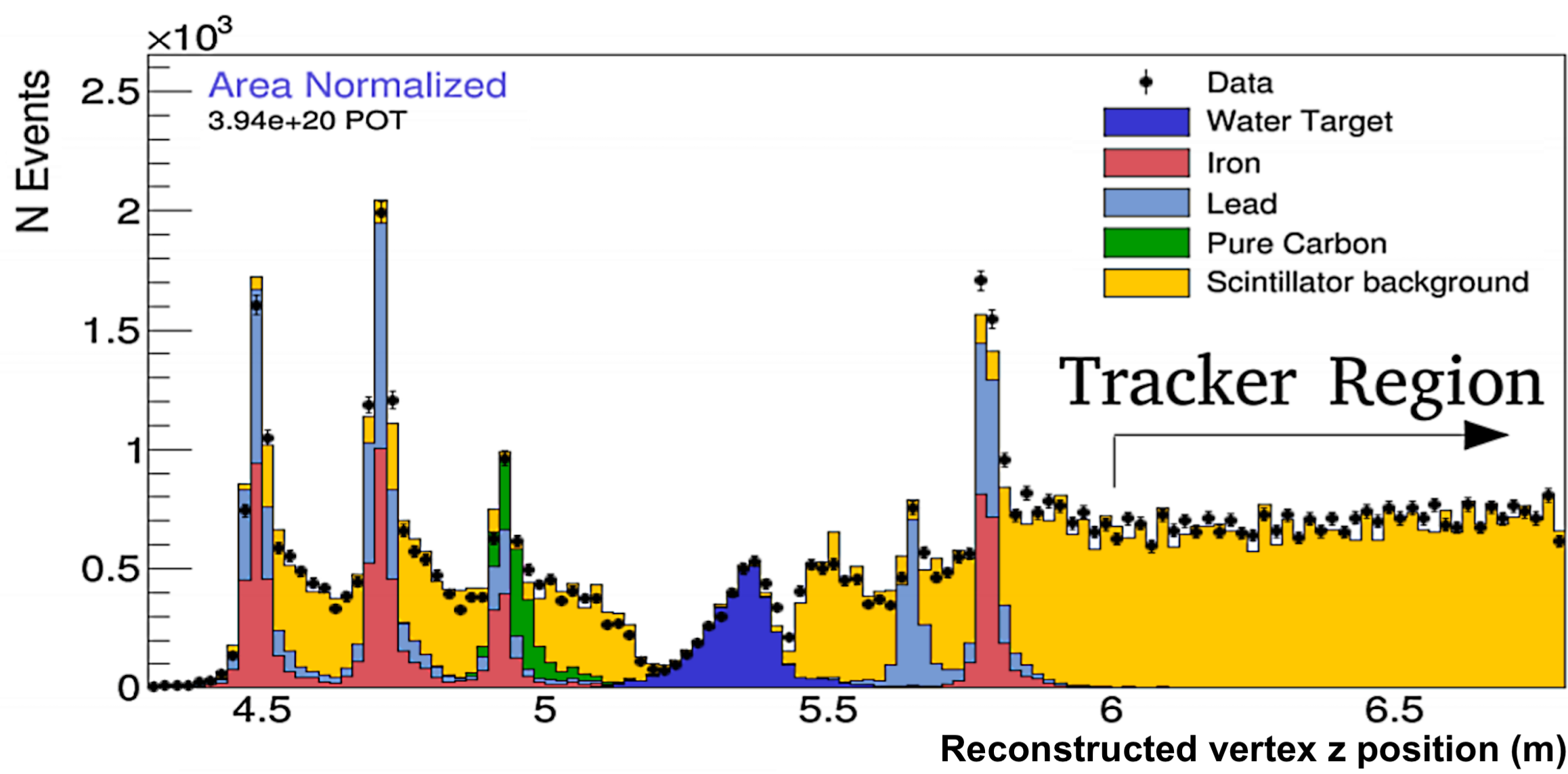}\\
    \caption{
    Reconstructed vertex location in the upstream region of MINERvA along the detector axis shown in data and simulation in the full water target configuration. Interactions correspond to those with two reconstructed tracks.    }
    \label{fig:zvertex}
\end{figure}

There are two primary categories of backgrounds:  interactions that originated in the scintillator but whose vertex is mis-reconstructed in a target, and non-quasielastic-like interactions that are correctly reconstructed in a nuclear target but are incorrectly reconstructed as quasielastic-like.  Predictions for both backgrounds are constrained by comparing the data to the simulation in sidebands.  The background between one nuclear target and another at the same vertex $z$ location is small due to fiducial volume cuts and is constrained by interactions in the other target. 

The prediction for the scintillator background can be constrained by the ratio between the data and the simulation for interactions reconstructed in the scintillator surrounding each of the nuclear targets.  Figure \ref{fig:zvertex} shows the reconstructed interaction vertex position as a function of distance along the detector axis for all interactions that have a reconstructed muon in MINOS and an additional reconstructed track:  the interactions in the denser nuclear target material show up as clear peaks in this distribution, and the normalization of the scintillator background comes from the interactions at least one scintillator plane away from each target.
\begin{figure}[h]
    \centering
\includegraphics[width=0.49\linewidth]{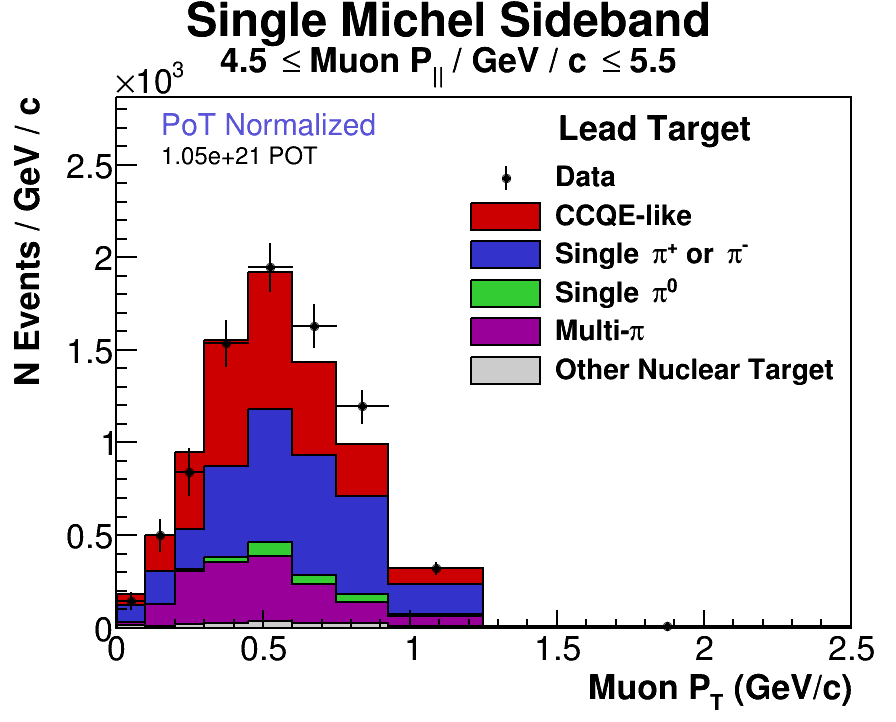}
\includegraphics[width=0.49\linewidth]{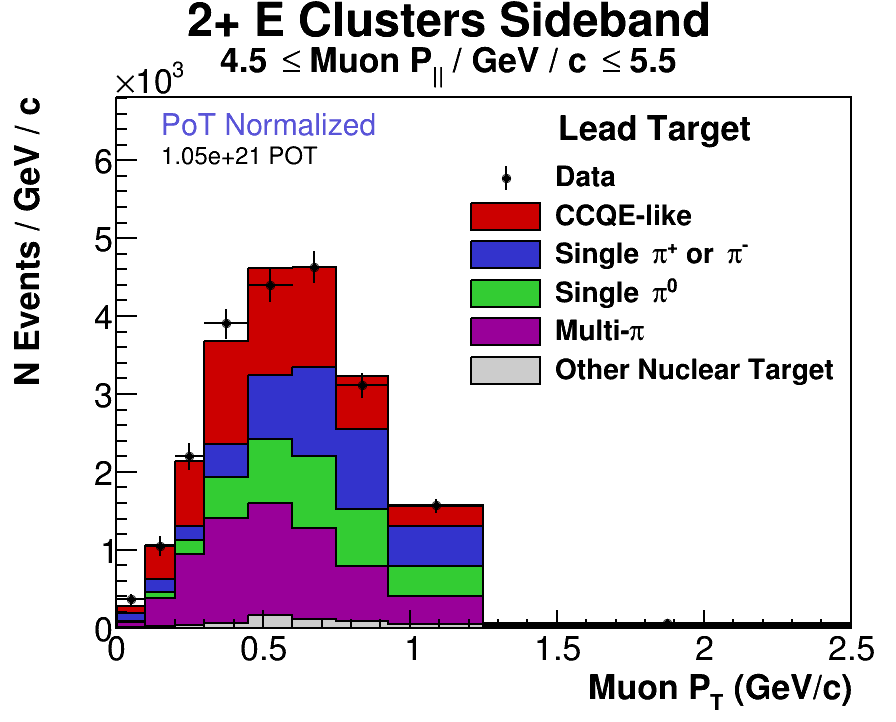}
\includegraphics[width=0.49\linewidth]{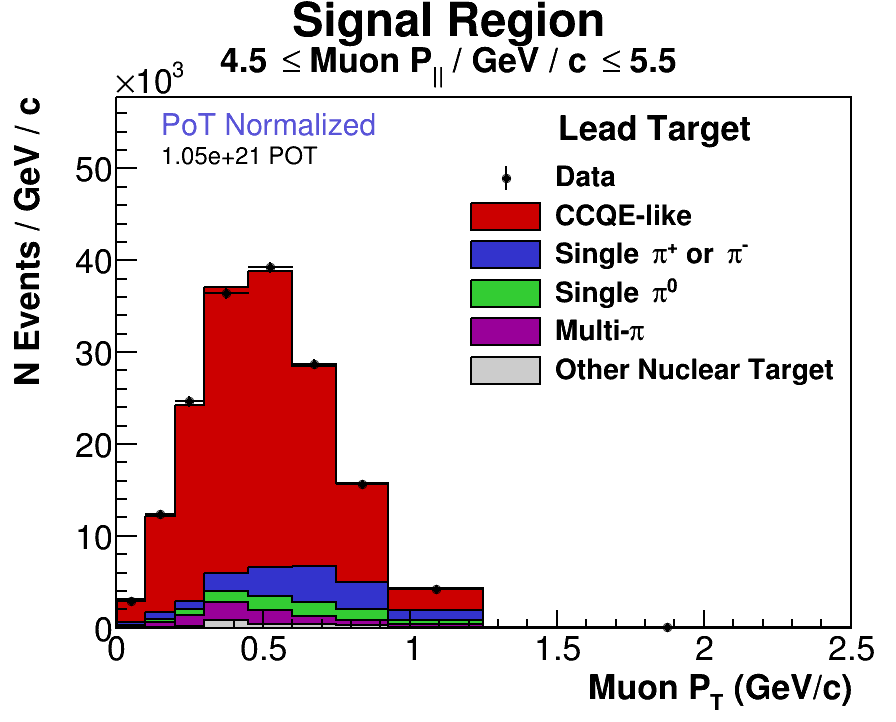}
\includegraphics[width=0.49\linewidth]{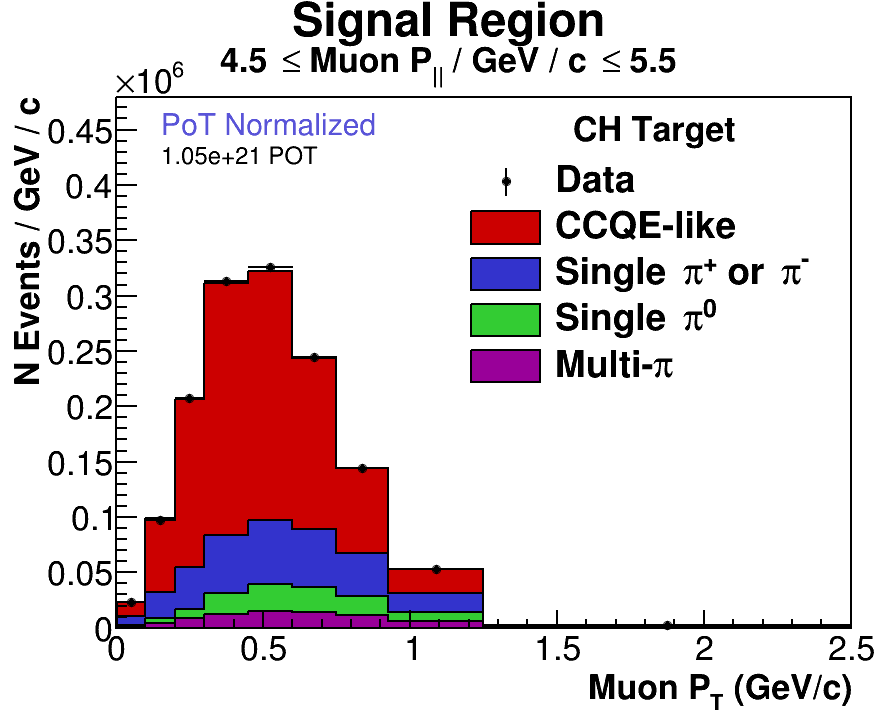}
    \caption{
   Top:  Data and prediction for the (left) single Michel Electron sideband, (right) Extra Energy cluster sideband for Pb. Bottom:  
  signal region in Pb (left) and CH (right), all after the backgrounds and the signal have been tuned, 
  for the peak longitudinal momentum (\pparmu)  bin. The scintillator background to Pb has been constrained and subtracted.  
   }
    \label{fig:phys_sidebands}
\end{figure}

The second category of backgrounds comes from interactions that take place 
in the target of interest, but 
are not quasielastic-like.  In this case one or more neutral or charged pions have been misidentified as a proton or not seen at all. The single neutral pion background is constrained by MINERvA's earlier measurement of neutral pion production~\cite{MINERvA:2020anu}.
To determine the backgrounds from other neutrino interaction channels, two different sidebands are used where the data is compared to the simulation.  The first sideband requires a Michel electron to provide a sample enriched with charged pions; the second requires at least two extra clusters of energy away from the interaction vertex to provide a sample enriched in neutral pions, as in Ref.~\cite{MINERvA:2019gsf}.

Figure \ref{fig:phys_sidebands} (top) shows the data and simulation in the two sidebands for the Pb target as a function of transverse momentum (\ptmu ) in the peak longitudinal momentum (\pparmu) region ($4.5<$\pparmu $/GeV/c<5.5$).  The top left plot shows the Michel electron sideband, and the center plot shows the interactions with two or more clusters of energy.  The plots on the bottom show the data and prediction with the signal prediction tuned to match the data in that region in the Pb (left) and scintillator (right).   The background levels from non-quasielastic-like interactions are $36\%$ in the scintillator and are $33-45\% $ in the nuclear targets, with the lowest background in the Pb targets.  The fact that the physics backgrounds are lower in Pb than in the lighter nuclei stems from the fact that pion production is suppressed in heavier nuclei relative to C, as measured by MINERvA~\cite{MINERvA:2022djk}.   

\begin{figure}[h]
    \centering
\includegraphics[width=\linewidth]{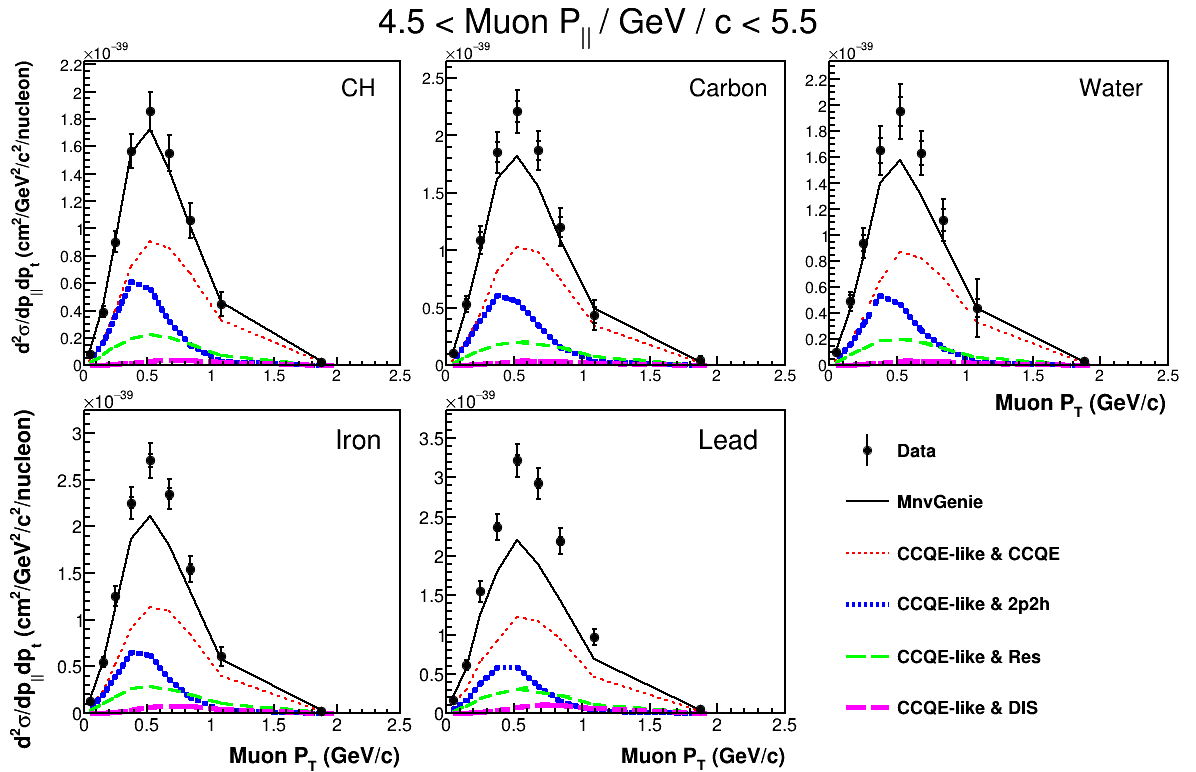}
    \caption{Cross section versus transverse muon momentum (\ptmu) in the highest statistics \pparmu\ bin in data and simulation
    for the nuclear target materials and for scintillator.  Inner (outer) error bars represent the statistical (total) uncertainties.
}
    \label{fig:xsec}
\end{figure}

\begin{figure}[tp]
    \centering
\adjustbox{trim={0.0\width} {.00\height} {0.0\width} {0.075\height},clip}%
{\includegraphics[width=.99\linewidth] {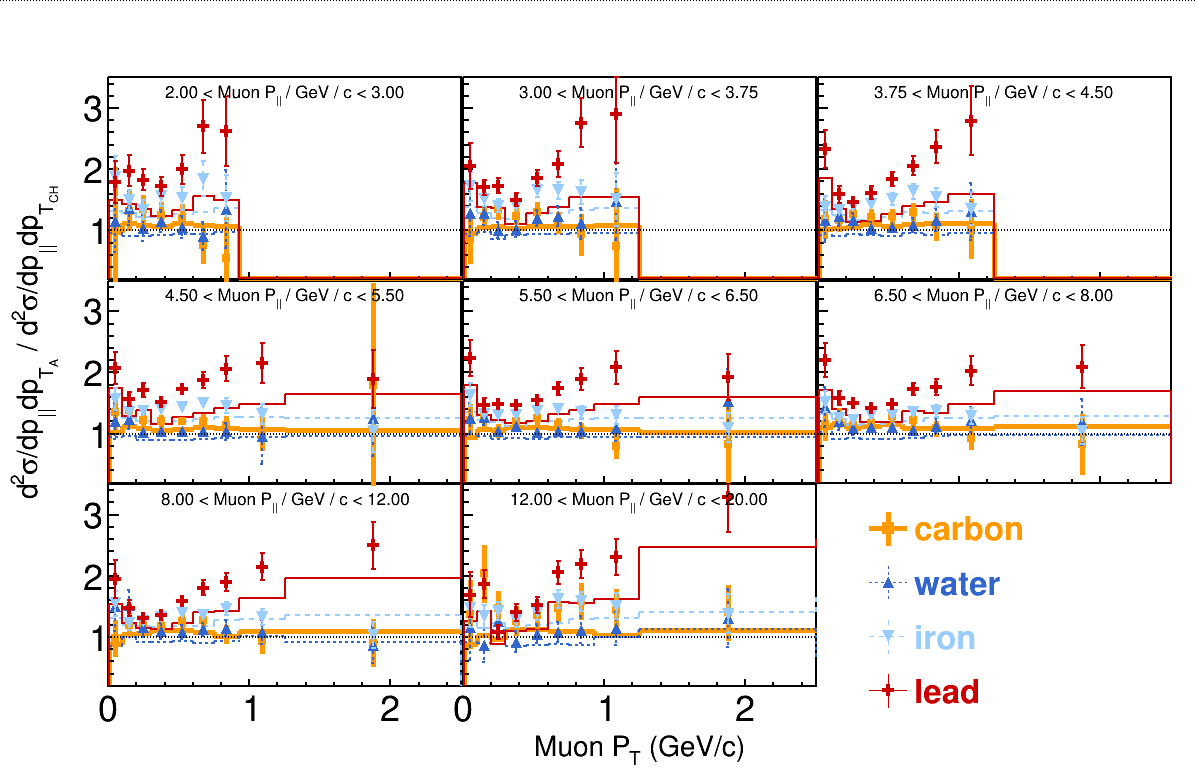}}
    \caption{Quasielastic-like cross-section ratios to scintillator versus muon momenta on Pb, Fe, water, and C.  The points are the data and the solid lines are the predictions from the model described in the text.}
    \label{fig:xsec_2d2}
\end{figure}

After background subtraction there are a million interactions in the scintillator tracker region, 25,000 interactions in the C target (used as a control region), and 20,000, 92,000, and 124,000 interactions in the water, Fe, and Pb targets, respectively.  
The analysis unfolds the distributions to correct for detector resolution using the D'Agostini prescription~\cite{DAgostini:1994fjx,DAgostini:2010hil}, and then corrects each of the different target regions for efficiency.  Finally, the cross section is found by dividing by the number of target nucleons and by the total integrated flux appropriate for each target.  Figure~\ref{fig:xsec} shows the cross section in data and simulation in all five target materials as a function of \ptmu\ for one \pparmu\  bin.  There is a clear excess above the prediction that grows as a function of the mass number $A$, and is consistent across \pparmu\  as shown in Fig.~\ref{fig:xsec_2d2}.  The prediction includes not only quasielastic and multi-nucleon interactions, but also interactions where some original final-state particles were absorbed in the nucleus.  The quasielastic interactions dominate at the highest \ptmu, but at low \ptmu\  there are significant contribution from both 2p2h interactions and pion production followed by absorption.  

The neutrino flux changes at the few per cent level as a function of position across the front face of the MINERvA detector\cite{MINERvA:2021mpk}. The nuclear targets do not all have the same integrated neutrino flux because each target covers only a part of the hexagonal shape of the scintillator planes.  In order to calculate the cross section on each nuclear target material, a different flux must be used for each material\cite{MINERvA:2022djk}.  The cross sections shown in Figure~\ref{fig:xsec} have been calculated using this prescription.  The systematic uncertainties in the absolute cross sections are dominated in most bins by the flux uncertainties. 

In order to minimize the flux uncertainties, the ratio of cross sections between a given target material and scintillator are reported where by construction the incident neutrino flux is the same to better than a per cent in both targets.  To do this, the analysis extracts the cross section in the scintillator in 12 different transverse wedges of the detector and then the scintillator cross section used in the ratio is the one calculated using the linear combination of wedges that most closely matches the illumination of each target material.  
%

Systematic uncertainties on the cross-section measurement arise from three different sources:  the flux, neutrino interactions, and the MINERvA detector (both the detector response and the target masses).  These uncertainties are evaluated using a multi-universe technique where the cross section is re-extracted after varying each source of uncertainty, and the correlations between different bins (and different nuclear targets) are taken into account.  
The flux uncertainty comes from uncertainty in hadron production and focusing effects, and is constrained at 3.9\% using neutrino-electron scattering interactions from the same exposure~\cite{MINERvA:2019mkf}. Neutrino interaction uncertainties are dominated by the modeling uncertainties in background processes, in particular the final-state interaction uncertainties.  Detector uncertainties are dominated by uncertainties in muon reconstruction, which are small, but increase at high \ptmu\ where the cross section is falling steeply and small changes in the muon energy scale have a large effect on the accepted interactions.  Because the systematic uncertainties are highly correlated between different targets there is significant reduction of the total uncertainty in the cross-section ratio measurements.  

The systematic uncertainties in the absolute cross sections, shown in the supplemental material, are dominated by the flux uncertainties and the muon energy scale.  Those uncertainties cancel to first order when taking ratios between the target material and the scintillator target, and the remaining largest systematic uncertainties come from the reconstruction uncertainties that do not cancel, for example those from final-state interactions in the target nuclei.  
  
The systematic and statistical uncertainties on the cross-section ratios, also shown in the supplemental material, are of comparable size in the most populated longitudinal momentum bin; in other bins the uncertainty is dominated by the statistical uncertainty.  Since most of the neutrino momentum is forward, the broad neutrino energy beam populates the \pparmu\  bins between $3.75$~GeV/c and $6.5$~GeV/c.  In most kinematic regions the cross-section ratio uncertainty is well below ten per cent.  

Figure~\ref{fig:xsec_2d2} shows the measurement and prediction for ratios of the cross sections per nucleon as a function of \ptmu\   for different \pparmu\  bins.  Each panel in the plot shows the ratio for Pb, Fe, water, and C compared to scintillator.  The ratios themselves grow as a function of mass number, as expected, since the higher $A$ nuclei have a higher neutron to nucleon ratio.  However, the discrepancy with the base model also grows as a function of mass number.  The cross-section ratio between Pb and CH changes dramatically as a function of \ptmu, and less dramatically as a function of \pparmu.  This indicates that the size of the nuclear effects varies more as a function of momentum transfer than as a function of neutrino energy.  
The cross-section ratio between Fe and CH appears flatter as a function of \ptmu\  and \pparmu, with a scaling per nucleon of about 1.4-1.5.  MINERvA's underlying model, which was not tuned to Fe or Pb data,  predicts a ratio that is closer to 1.2. 

The discrepancy between data and simulation at high \ptmu\ implies that the total quasielastic-like cross-section scaling versus $A$ is higher than modeled, and that effect increases with increasing momentum transfer. 
The discrepancy at low \ptmu\  does not appear to grow with \pparmu; this implies that the $A$-dependence of interactions coming from 2p2h and/or pion absorption is underpredicted, although there is not a strong energy dependence.
The $A$-scaling for single pion production on Fe and Pb has been measured to be lower than predicted~\cite{MINERvA:2022djk}.  This could come from more pion absorption than current models predict, which would then present as higher $A$-scaling for quasielastic-like interactions that result from pion absorption.  The cross-section ratios between water and scintillator appear to be consistent with unity with no significant dependence on the muon kinematics seen at the 10\% level, as shown in Fig.~\ref{fig:xsec_2d2}.  

\begin{figure}[h]
    \centering
\adjustbox{trim={0.0\width} {.00\height} {0.0\width} {0.088\height},clip}%
{\includegraphics[width=.99\linewidth] {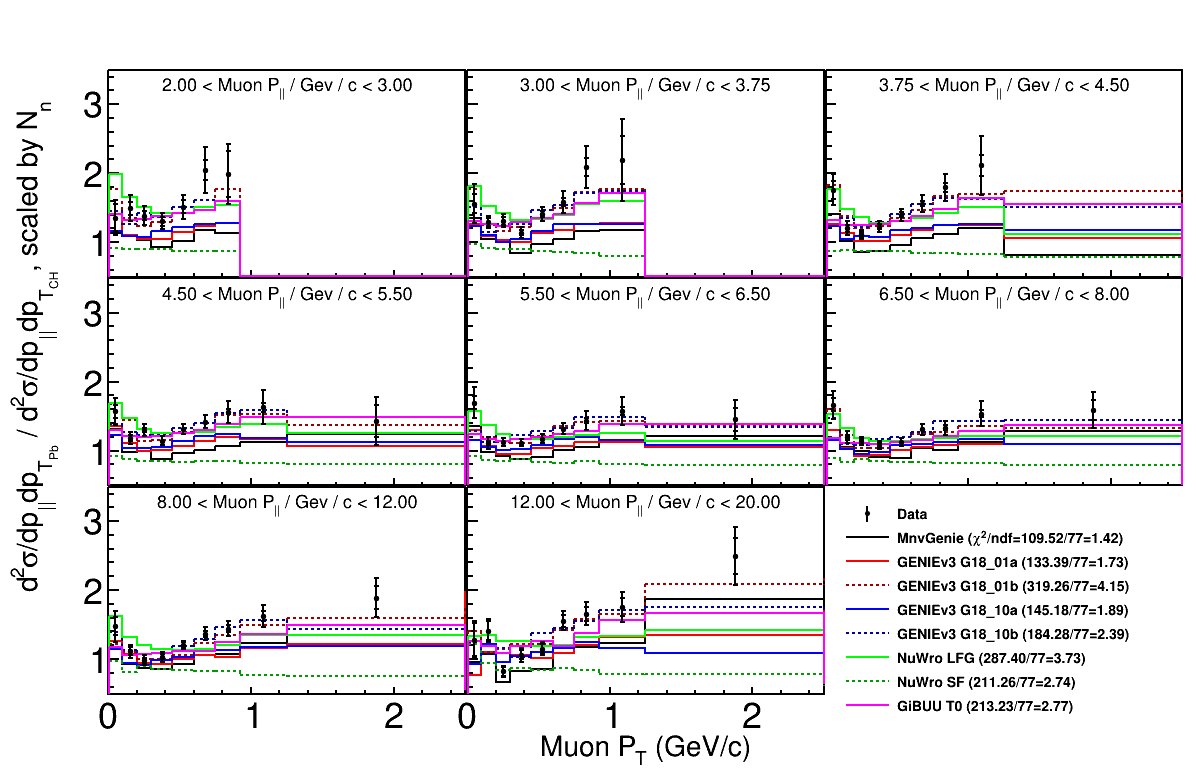}}
    \caption{Comparison between several models for quasielastic-like scattering and the data on the Pb to CH cross-section ratio, along with the $\chi^2$ between each model and the data. 
    }
    \label{fig:models}
\end{figure}

Figure~\ref{fig:models} shows the cross-section ratios for Pb/CH compared to different model choices in GENIE and NuWro~\cite{NUWRO}.  A comparison to GIBUU~\cite{GIBUU} which uses a microscopic cascade model to describe final-state interactions is also shown. 
None of the generator predictions are in good agreement with the data.  Different final-state-interaction models in GENIE change in the cross-section ratio prediction, especially at high \ptmu.  The data prefer GENIE’s hA model which approximates intranuclear rescattering as a single effective interaction within the nucleus, to its hN model which is a microscopic cascade model.
However the overall performance of GIBUU in Fig.~\ref{fig:models} may indicate that models of the latter type are better suited to characterize pion intranuclear absorption in heavy nuclei. 

The difference in $A$-scaling that arises between using the Relativistic Fermi Gas with the Bodek-Ritchie tail (BRRFG) and the Local Fermi Gas (LFG) initial-state nucleon models is much smaller than what arises from different final-state interaction models in GENIE.  This may be because the choice of BRRFG or LFG only affects the quasielastic process and not 2p2h or resonance production. 
Changing the initial nucleon state makes a larger change in the NuWro model, where the data prefer the Spectral Function (SF) over the LFG treatment, although neither agrees as well with the data
as the GENIE hA models.  

MINERvA has measured quasielastic-like cross-section ratios and sees evidence of scaling as a function of $A$ that is not constant over the momentum transferred to the nucleus, and not predicted by any generators considered.  MINERvA's measurement of pion production on these same nuclei~\cite{MINERvA:2022djk} implies that for higher $A$ nuclei, more pions are being absorbed compared to what one would predict given the pion production measured on CH. These measurements combined provide key benchmarks for the field's description of how the nucleus impacts neutrino interactions. 

This document was prepared by members of the MINERvA Collaboration using the resources of the Fermi National Accelerator Laboratory (Fermilab), a U.S. Department of Energy, Office of Science, HEP User Facility. Fermilab is managed by Fermi Research Alliance, LLC (FRA), acting under Contract No. DE-AC02-07CH11359. These resources included support for the MINERvA construction project, and support for construction also was granted by the United States National Science Foundation under Award No. PHY-0619727 and by the University of Rochester.  Support for  participating scientists was provided by NSF and DOE (USA); by NSERC (Canada); by CAPES and CNPq (Brazil); by CoNaCyT (Mexico); by Proyecto Basal FB 0821, CONICYT PIA ACT1413, and Fondecyt 3170845 and 11130133 (Chile);  by CONCYTEC (Consejo Nacional de Ciencia, Tecnolog\'ia e Innovaci\'on Tecnol\'ogica), DGI-PUCP (Direcci\'on de Gesti\'on de la Investigaci\'on  - Pontificia Universidad Cat\'olica del Peru), and VRI-UNI (Vice-Rectorate for Research of National University of Engineering) (Peru); NCN Opus Grant No. 2016/21/B/ST2/01092 (Poland); by Science and Technology Facilities Council (UK); by EU Horizon 2020 Marie Skłodowska-Curie Action.  We thank the MINOS Collaboration for use of its near detector data. Finally, we thank the staff of Fermilab for support of the beam line, the detector, and computing infrastructure.

\clearpage
\onecolumngrid
\pagebreak

\appendix{Supplemental Material}
\section{ Uncertainties on cross section ratios} 
\begin{figure*}[h]
    \centering
\includegraphics[width=\linewidth] 
{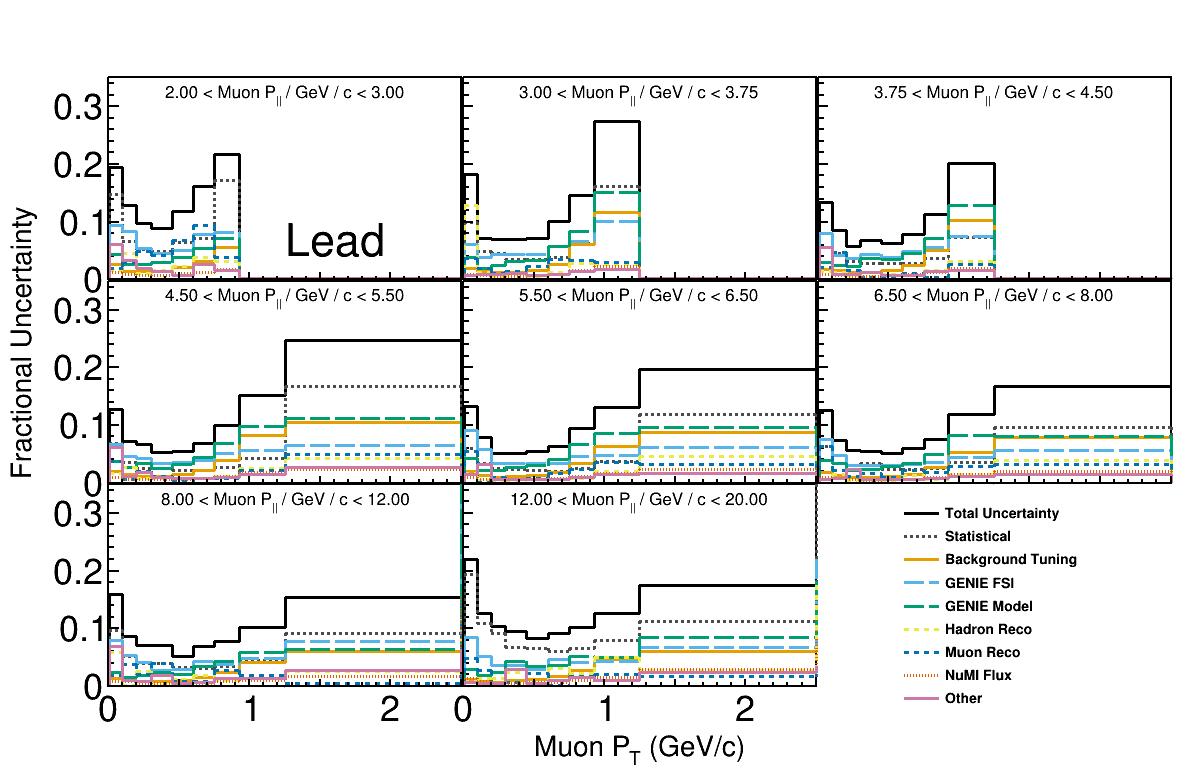}
    \caption{Uncertainties on the cross section ratio of Lead to Scintillator as a function of muon kinematics.  }
    \label{fig:syst_err}
\end{figure*}
\newpage
\section{Uncertainties on Cross Sections on Pb and CH}
\begin{figure*}[h]
    \centering
    \includegraphics[width=.56\linewidth]{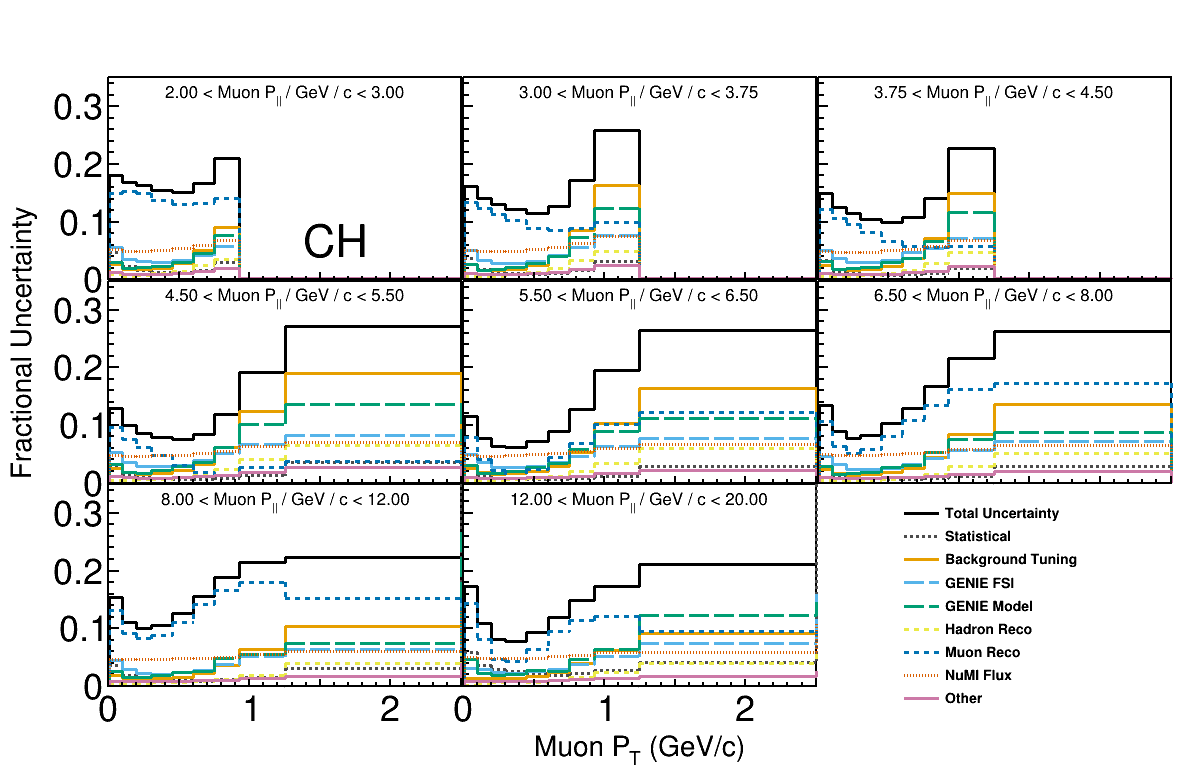}  \includegraphics[width= 0.56\linewidth]{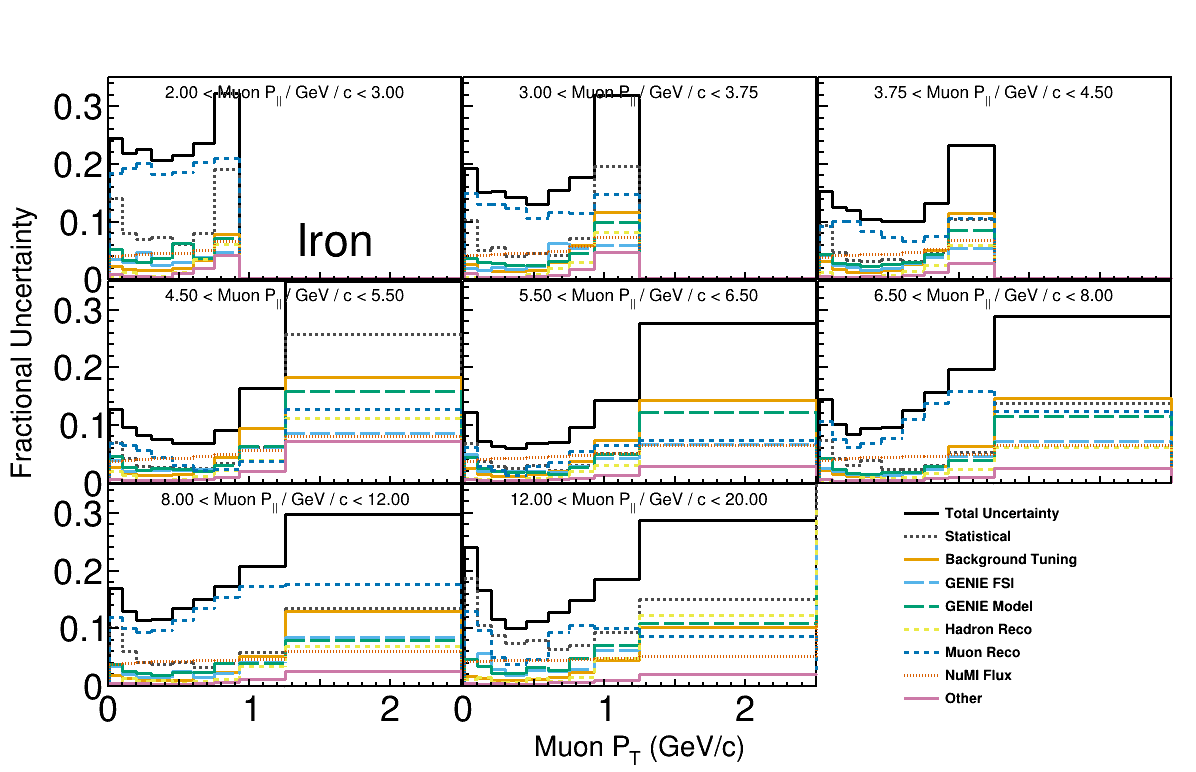}
\includegraphics[width=.56\linewidth]{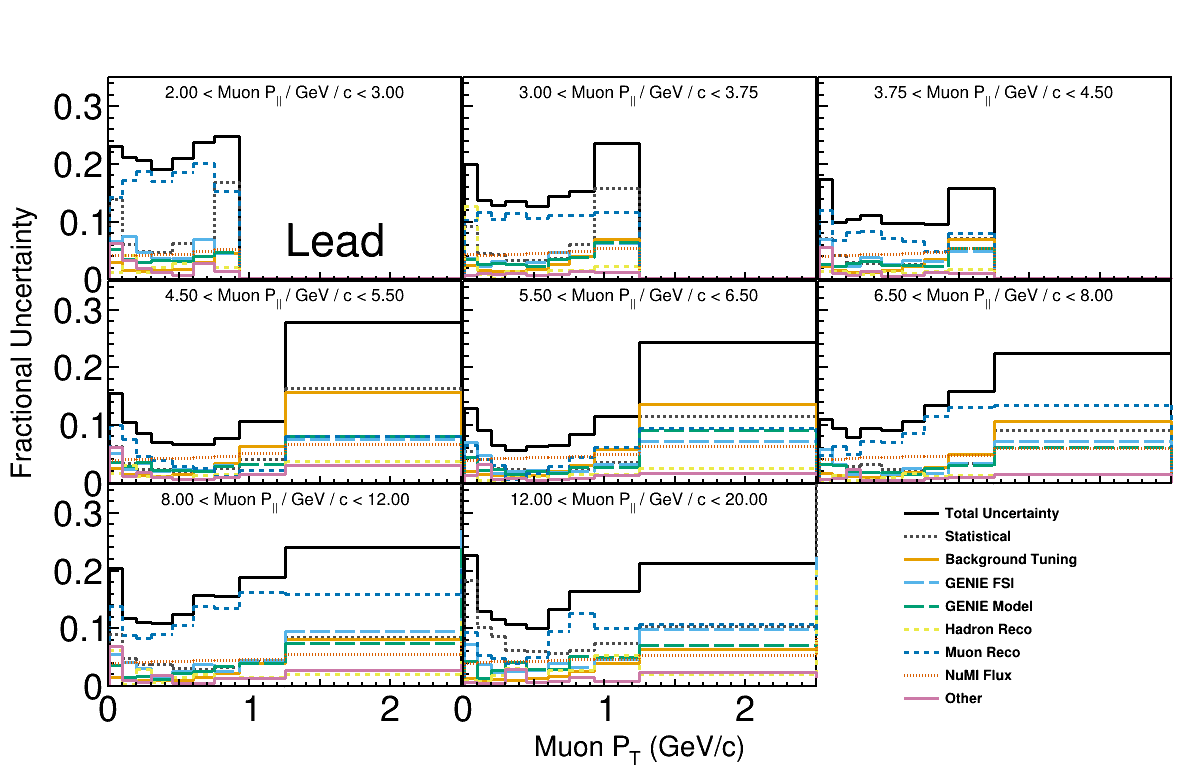}
    \caption{Cross Section Uncertainties on the scintillator tracker (top) and Pb (bottom) as a function of longitudinal and transverse momentum}
    \label{fig:sup_xsec_uncertainties}
\end{figure*}

\newpage
\section{Cross Sections on iron, lead and CH}
\begin{figure*}[h]
    \centering
    \includegraphics[width= 0.56\linewidth]{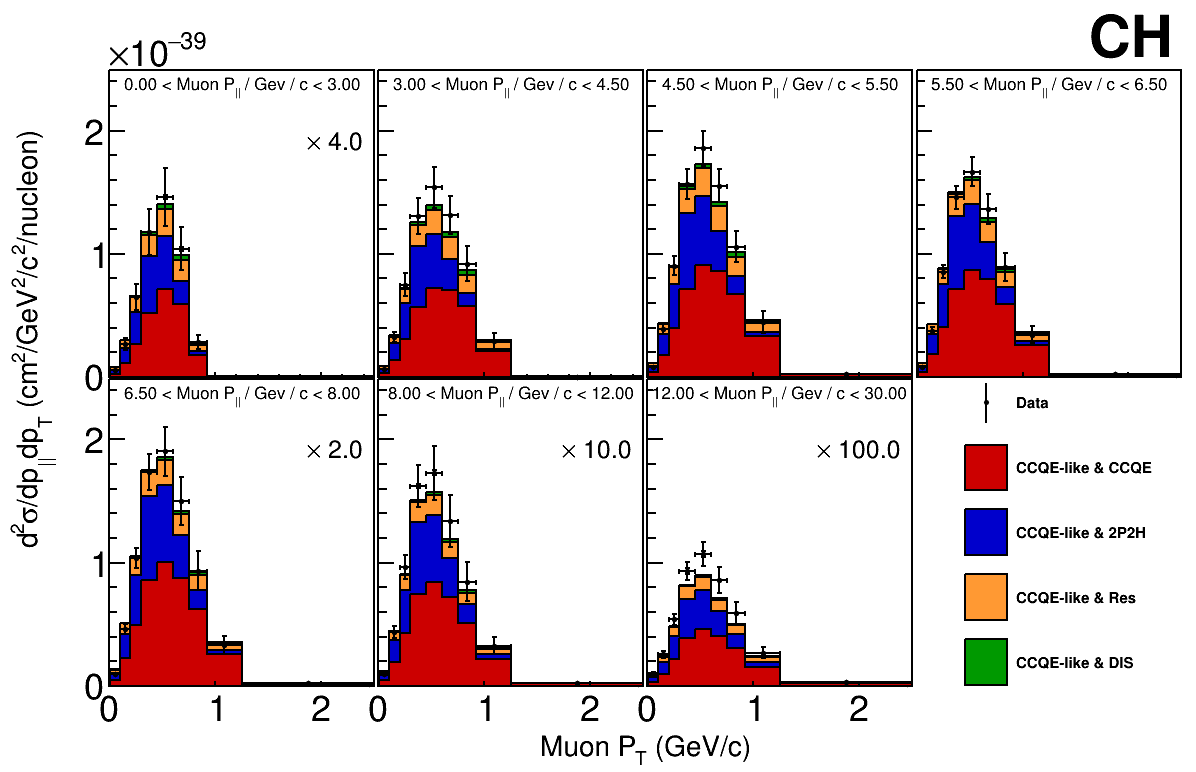}
    \includegraphics[width= 0.56\linewidth]{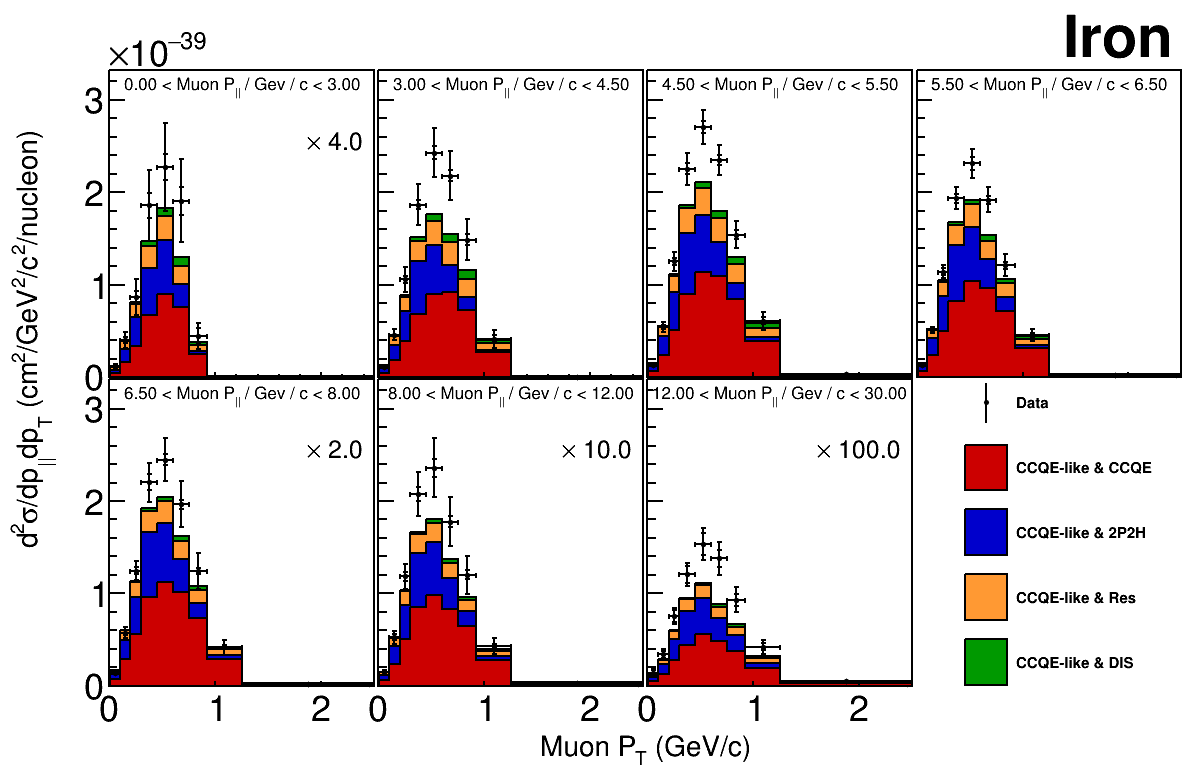}
    \includegraphics[width= 0.56\linewidth]{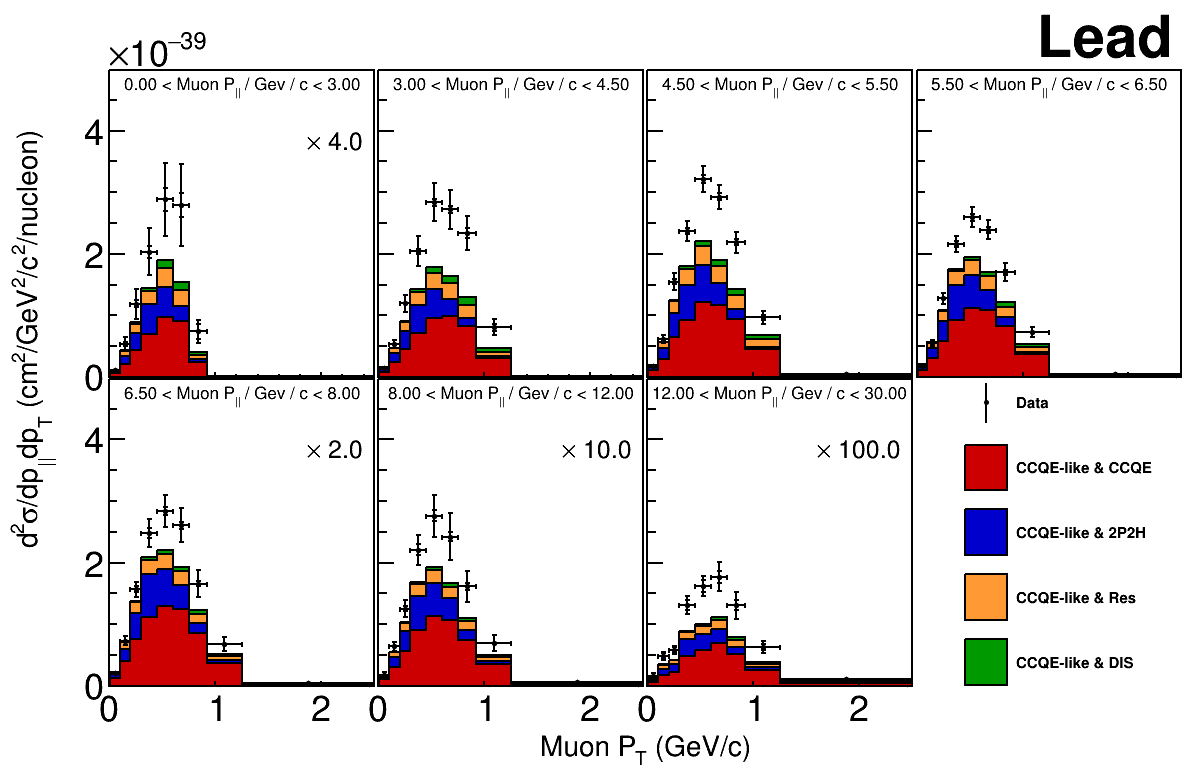}
    \caption{Cross Section on the scintillator tracker (top), iron (middle) and lead (bottom) as a function of longitudinal and transverse momentum}
    \label{fig:sup_xsec_ch_2d}
\end{figure*}
\end{document}